\documentstyle[11pt,newpasp,twoside,psfig]{article}

\index{summary}
\index{instructions}
\index{template}

\def\edcomment#1{\iffalse\marginpar{\raggedright\sl#1\/}\else\relax\fi}
\reversemarginpar

\newcommand{\be}{\begin{equation}}
\newcommand{\ee}{\end{equation}}

\newcommand{\ba}{\begin{eqnarray}}
\newcommand{\ea}{\end{eqnarray}}

\begin{document}
\title{
Magnetospheric Structure and Non-Thermal Emission of AXPs and SGRs}

\author{
M. Lyutikov\altaffilmark{1,2,3},
C. Thompson\altaffilmark{4},
S.R. Kulkarni\altaffilmark{5}
}
\vskip 1in
\altaffiltext{1}{Department of Physics, McGill University,
Montr\'eal, QC}
\altaffiltext{2}{Massachusetts Institute of Technology,
77 Massachusetts Avenue, Cambridge, MA 02139}
\altaffiltext{3} {CITA National Fellow}
\altaffiltext{4}{Canadian Institute for Theoretical Astrophysics,
60 St. George St., Toronto, ON M5S 3H8}
\altaffiltext{5}{California Institute of Technology, 105-24, Pasadena,
CA 91125}

\begin{abstract}
In the framework of the magnetar model for the Soft Gamma Repeaters 
and Anomalous X-ray Pulsars, we consider the structure of neutron 
star magnetospheres threaded by large-scale electrical currents.
We construct self-similar, force-free equilibria under the assumption
of axisymmetry and a power law dependence of magnetic field on radius, 
${\bf B} \propto r^{-(2+p)}$.  A large-scale twist of the field lines 
softens the radial dependence to $p < 1$, thereby accelerating the
spindown torque with respect to a vacuum dipole.   A magnetosphere 
with a strong twist ($B_\phi/B_\theta = O(1)$ at the equator) has an 
optical depth $\sim 1$ to resonant cyclotron scattering,
independent of frequency (radius), surface magnetic field strength, or the
charge/mass ratio of the scattering charge.  We investigate the effects of the 
resonant Compton scattering by the charge carriers (both electrons and ions)
on the emergent X-ray spectra and pulse profiles.  
\end{abstract}

\section{Nature of the AXPs and SGRs}

The Anomalous X-ray Pulsars and Soft Gamma Repeaters are neutron stars
which share similar spin periods, $P = 5 -- 12$ s, characteristic ages,
$P/\dot P = 3\times 10^3 - 4\times 10^5$ yr, and X-ray luminosities,
$L_X = 3\times 10^{34} - 10^{36}$ erg s$^{-1}$, well in excess
of the spin-down luminosities (Thompson et al. 2001).
They are almost certainly young and isolated:
no evidence for a binary stellar companion has yet been detected in 
any of these sources, and a few are convincingly associated with 
young supernova remnants.   The overlap between the SGR and AXP sources 
in a three-dimensional parameter space ($P$, $\dot{P}$ and $L_X$), 
and the observed variability in their X-ray output, provides 
circumstantial evidence that they share a common energy source:  
the decay of a very strong ($\ga 10^{15}$~G) magnetic field.

The persistent emission of the SGRs and AXPs has both a thermal
and a non-thermal component.  In the case of the SGRs, this emission
becomes brighter after periods of bursting activity 
and involves a comparable release of energy to the outbursts 
(averaged over time).

\section{Twisted Neutron Star Magnetospheres}

The magnetic fields of magnetars are most likely generated by
a hydromagnetic dynamo as the star is born, and may be associated
with rapid initial rotation (Duncan \& Thompson 1992).  A strong
twist in a $\sim 10^{15}$ G magnetic field will relax at intervals 
as the field is transported through the deep interior of the neutron star.
Even if the electrical current were initially confined to interior
of the star, the Lorentz force would become strong enough to deform its
crust, thereby {\it twisting up} the external magnetic field.   In 
such a situation, a persistent current will flow through the 
magnetosphere, supported by emission of light ions (e.g. H, He, C)
 and electrons 
from the neutron star surface.  The decay of this current outside of 
the star is an efficient mechanism for converting magnetic energy 
to X-rays, and for inducing rapid variations in the X-ray flux.
 
To see how the persistent current will modify the structure of the 
magnetosphere, we find axisymmetric force-free equilibria outside
a (non-rotating) spherical surface,
${\bf\nabla}\times{\bf B} = \alpha({\cal P}) {\bf B}$.  These equilibria
form a one-dimensional sequence labeled by 
the flux parameter ${\cal P} = {\cal P}(R,\theta)$, with poloidal
magnetic field
$
{\bf B}_P = {{\bf\nabla}{\cal P} \times {\hat\phi}/ R\sin\theta}.
$
As a major simplification we consider self-similar configurations 
\be {\cal P} = {\cal P}_0 (R/R_{\rm NS})^{-p}F(\theta), 
\;\;\;\;\;\; B(R,\theta) \sim F(\theta) \times  (R/R_{\rm NS})^{-(2+p)},
\ee
following Lynden-Bell \& Boily (1994).
The radial index $p$ is uniquely determined by a single parameter $C$,
which is related to the strength of the current:
\be
p(p+1)F + \sin^2\theta{\partial^2 F\over\partial(\cos\theta)^2} = -CF^{1+2/p}.
\ee
Choosing a dipole field at a zero current, $p(C)$ is determined by the
{\it three} boundary conditions, 
$ B_r (R,\theta=\pi/2) =0, \,
B_r (R_{\rm NS}, \theta=0)  = {\rm const} , \,
B_{\phi} (R,\theta=0)  = 0.
$
The index $p$ 
is most conviently expressed as a function of the net twist 
$\Delta\phi_{\rm N-S}$ between the north and south magnetic poles (Fig. 1a).
\begin{figure}
\centerline{
\psfig{file=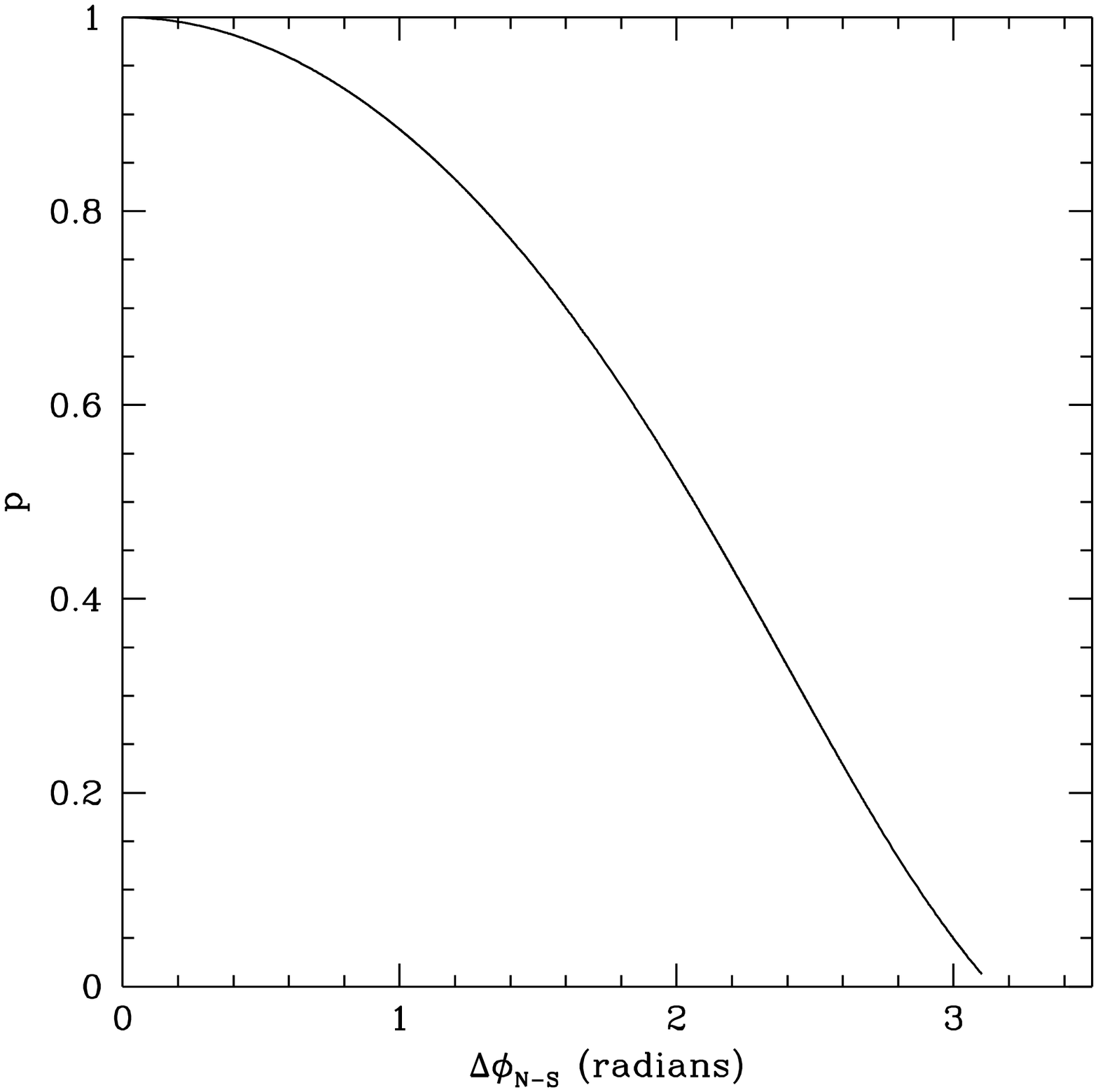,width=5cm}
\psfig{file=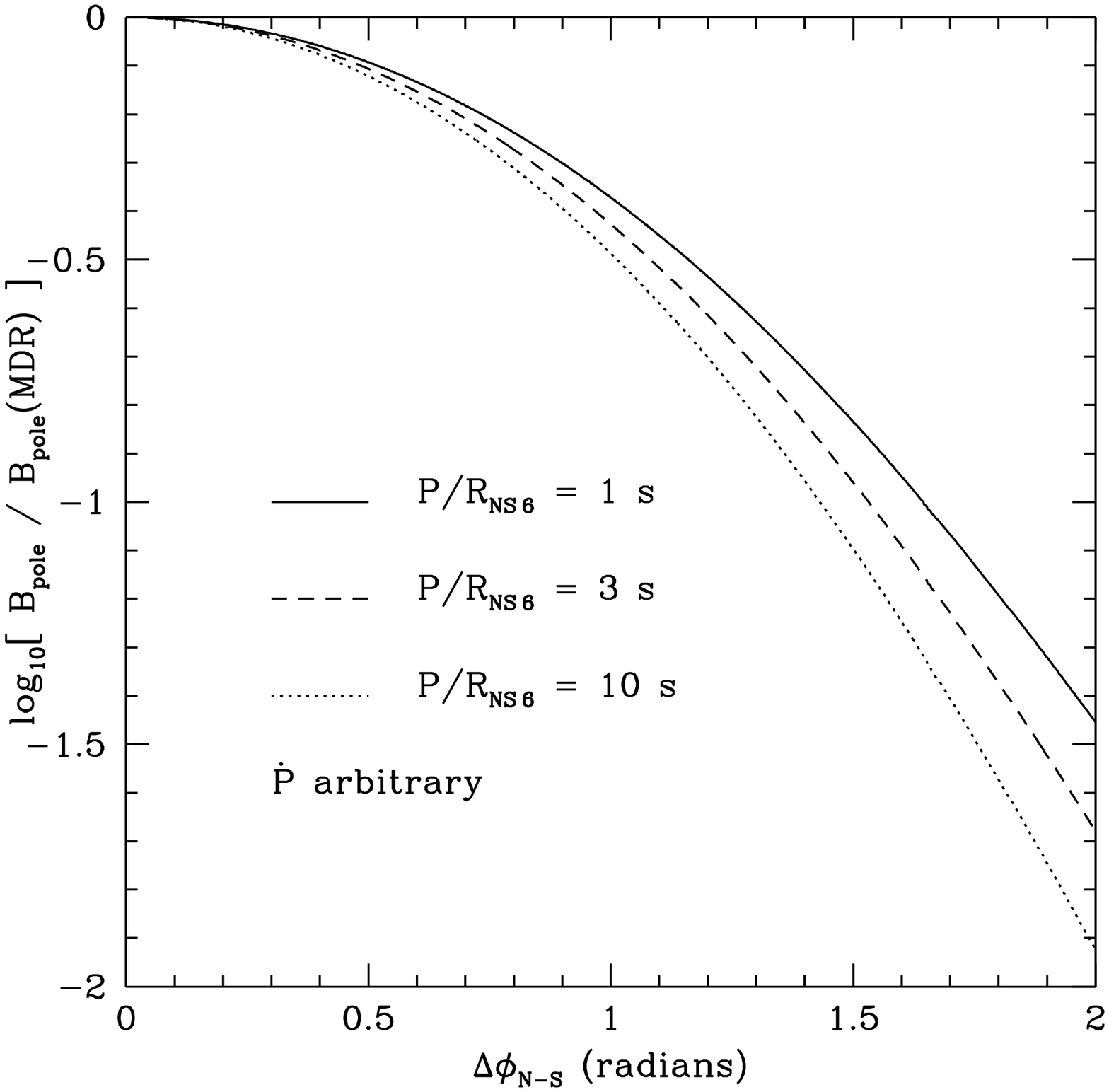,width=5cm}}
\caption{(a)  The radial index $p$ as a function of the
twist $\Delta\phi_{\rm N-S}$.
(b) The actual polar field inferred from spin parameters $P$ and $\dot P$,
compared with the magnetic dipole formula.}
\end{figure}

\section{Resonant  Scattering}

The current-currying charges also provide a significant optical 
depth to resonant cyclotron scattering.  For a particle of charge $Ze$ 
and mass $M$, the resonant cross-section is
$
\sigma_{\rm res}(\omega) = (\pi^2 Ze^2/Mc)\,(1+\cos^2\theta_{kB})
\delta(\omega-\omega_c)$.  The optical depth is
determined by relating the particle density $n_Z$ to the twist in the
magnetic field
through $(Ze) n_Z v_Z = \epsilon_Z(c/4\pi)|{\bf \nabla}\times{\bf B}|$.
In our self-similar model,
\be
\left({v_Z\over c}\right)\,\tau_{\rm res} =
{\pi\epsilon_Z\over 8}\,\left(1+\cos^2\theta_{kB}\right)\,
\left[{F(\theta)\over F(\pi/2)}\right]^{1/p}\,\Delta\phi_{\rm N-S}.
\ee
Thus the optical depth is 
 $ \tau_{\rm res} \sim (v_Z/c)^{-1}$ for strong twists  
$\Delta\phi_{\rm N-S} \sim 1$.  Remarkably it is {\it 
independent of the mass and charge of the scatterers, the radius,
or the resonant frequency}.

\section{Implications for X-ray Spectra and Pulse Profiles}
\begin{itemize}
\item{ \it Surface Heating.} 
If ions and electrons supply the current, then the stellar surface
is heated at the rate
\be
L_X  
\simeq 3\times 10^{35}\,\epsilon_{\rm ion}
\left({B_{\rm pole}\over 10^{14}~{\rm G}}\right)\,
\left({B_\phi\over B_\theta}\right)_{\theta = \pi/2}
\hskip .3 truein {\rm erg~s^{-1}}.
\ee
(The electrons are electrostatically accelerated downward above the anode.)
This is comparable to the observed luminosities of AXPs and SGRs.
A global twist will decay on the timescale
\be
t_{\rm decay} = {E_B-E_B({\rm dipole})\over L_X}
= 30\,\epsilon_{\rm ion}^{-1}\left({B_{\rm pole}\over 10^{14}~{\rm G}}\right)\,
\Delta\phi_{\rm N-S}\hskip .3 truein {\rm yr}.
\ee
\item{\it Resonant Comptonization.}
Both ions and electrons can be expected to move mildly relativistically
where they resonantly scatter $1-10$ keV photons.  The product of 
the mean frequency shift $\langle\Delta\omega/\omega\rangle$ per scattering 
with the expected number of scatterings is $O(1)$ when $\Delta\phi_{\rm N-S} 
\sim 1$ radian.  Therefore, multiple scattering at the
cyclotron resonance by the moving charge carriers
will create a non-thermal spectral tail to the X-ray flux
emerging from the surface.  At optical depths $\la 1$, the hardness of 
the spectrum increases with the number of scattering and thus with the 
resonant optical depth. 
\item{\it Pulse Profiles.}
The emergent pulse profile is strongly modified by magnetospheric 
scattering.  Three effects enter here:  the optical depth 
$\tau_{\rm res}(\theta)$ is anisotropic, vanishing toward the magnetic
poles;  the resonant surface is aphserical;  and the scattered
radiation tends to be beamed along the magnetic field (in part
because of the motion of the charge carriers).  The pulse profile
will be approximately frequency-independent in this self-similar model.
\item{\it Ion Cyclotron Resonance.}
The ion component of the current will generate a
{\it comparable} optical depth to the electron component in a self-similar
magnetosphere (at frequencies below the surface cyclotron frequency). 
Since the ion cyclotron resonance sits much closer to the star 
(at $\sim 10-20$ km for $2-10$ keV photons), it is more sensitive to 
the presence of higher magnetic multiples.  Thus, a combination of ion 
and electron scattering in a multipolar magnetic field will produce 
an energy dependent total profile which is more complex at higher energies.
An ion cyclotron emission line would be a clear observational signature 
of surface heating by magnetospheric charges.
\item{ \it SGR/AXP spindown.}
When twisted, the magnetic field drops off more slowly than $\sim R^{-3}$. 
Thus the real polar field inferred from $P, \dot{P}$ is smaller than
the magnetic dipole formula would imply, while the braking index
becomes $ n= 2p+1 < 3$ (Fig. 1b).
 Our model predicts  that the spindown rate increases
with the optical depth to resonant scattering, and hence with the
hardness of the persistent X-ray spectrum.
In fact the active SGRs
1806-20 and 1900+14 both have higher $\dot P$ and harder X-ray spectra
than any AXP.  The quiescent SGR 0525-66 has a softer
spectrum.  We predict that its spindown rate,
when measured, will be intermediate between these sources and the AXPs.  
\item{\it Giant Flare Mechanism}
There are two generic possibilities for the production of the
giant flares of the SGRs, in the framework of our model:
(i)  giant flares may result from a sudden change (unwinding) in the
internal magnetic field,  implanting a twist into
the magnetosphere; or alternatively (ii) the twist may build up
more gradually in the magnetosphere, leading to a sudden relaxation
in close analogy with Solar flares.  Measurements of hardening/softening
of the persistent X-ray spectrum, changes in pulse profile, and
changes in spindown rate provide a way of discriminating between
these possibilities.
  In particular, the simplified pulse profile
observed in SGR 1900+14 after the 27 August 1998 giant flare 
(Woods et al. 2000) can be explained by an increase in the currrent at the 
radius of the electron cyclotron resonance.

\end{itemize}


\begin{thebibliography}
\bibitem{tm} {{Duncan}, R.C. and {Thompson}, C.}, 1992, {\apjl}, 392, L9
\bibitem{lbb} {{Lynden-Bell}, D. and {Boily}, C.}, 1994, {\mnras}, 267, 146
\bibitem{tlk01} Thompson, C.,  Lyutikov M., Kulkarni S., {\tt astro-ph/0110677}
\bibitem{woods01} Woods, P., et al. 2001, {\apj}, 552, 748
\end{thebibliography}
\end{document}